\documentclass[prb,preprint,aps,superscriptaddress]{revtex4}

\usepackage{graphicx}
\usepackage{color,soul}
\usepackage{url}

\usepackage[version=3]{mhchem} 
\usepackage[T1]{fontenc}       

\begin{document}

\title{Large-Gap Two-Dimensional Topological Insulator in Oxygen Functionalized MXene}

\author{Hongming Weng}

\email{hmweng@iphy.ac.cn}

\affiliation{Beijing National Laboratory for Condensed Matter Physics,
  and Institute of Physics, Chinese Academy of Sciences, Beijing
  100190, China}

\affiliation{Collaborative Innovation Center of Quantum Matter,
  Beijing, China}

\author{Ahmad Ranjbar}
\affiliation{Computational Materials Science Research Team, RIKEN Advanced Institute for Computational Science (AICS), Kobe, Hyogo 650-0047, Japan}

\author{Yunye Liang}

\affiliation{New Industry Creation Hatchery Center, Tohoku University,
  Sendai, 980-8579, Japan}

\author{Zhida Song}

\affiliation{Beijing National Laboratory for Condensed Matter Physics,
  and Institute of Physics, Chinese Academy of Sciences, Beijing
  100190, China}

\author{Mohammad Khazaei}
\affiliation{Computational Materials Science Unit, National Institute for Materials Science (NIMS), 
1-1 Namiki, Tsukuba 305-0044, Ibaraki, Japan}

\author{Seiji Yunoki}
\affiliation{Computational Materials Science Research Team, RIKEN Advanced Institute for Computational Science (AICS), Kobe, Hyogo 650-0047, Japan}
\affiliation{Computational Condensed Matter Physics Laboratory, RIKEN, Wako, Saitama 351-0198, Japan}
\affiliation{Computational Quantum Matter Research Team, RIKEN Center for Emergent Matter Science (CEMS), Wako, Saitama 351-0198, Japan }

\author{Masao Arai}
\affiliation{Computational Materials Science Unit, National Institute for Materials Science (NIMS), 
1-1 Namiki, Tsukuba 305-0044, Ibaraki, Japan}

\author{Yoshiyuki Kawazoe}

\affiliation{New Industry Creation Hatchery Center, Tohoku University,
  Sendai, 980-8579, Japan}

\affiliation{Thermophysics Institute, Siberian Branch, Russian Academy
  of Sciences, Russia}

\author{Zhong Fang}
\affiliation{Beijing National Laboratory for Condensed Matter Physics,
  and Institute of Physics, Chinese Academy of Sciences, Beijing
  100190, China}

\affiliation{Collaborative Innovation Center of Quantum Matter,
  Beijing, China}

\author{Xi Dai}


\affiliation{Beijing National Laboratory for Condensed Matter Physics,
  and Institute of Physics, Chinese Academy of Sciences, Beijing
  100190, China}

\affiliation{Collaborative Innovation Center of Quantum Matter, Beijing, China}

\date{\today}

\begin{abstract}
Two-dimensional (2D) topological insulator (TI) have been recognized as a new class 
of quantum state of matter. They are distinguished from normal 2D insulators with their nontrivial 
band-structure topology identified by the $Z_2$ number as protected by time-reversal symmetry (TRS). 2D TIs
have intriguing spin-velocity locked conducting edge states and insulating properties in the bulk. In the edge states,
the electrons with opposite spins propagate in opposite directions and the backscattering is fully 
prohibited when the TRS is conserved. This leads to quantized dissipationless ``two-lane highway" for charge 
and spin transportation and promises potential applications. Up to now, only very few 2D systems have
been discovered to possess this property. The lack of suitable material 
obstructs the further study and application. Here, by using first-principles calculations, we propose that the 
functionalized MXene with oxygen, M$_2$CO$_2$ (M=W, Mo and Cr), are 2D TIs with the largest gap of 
0.194 eV in W case. They are dynamically stable and natively antioxidant. Most importantly, they are very 
likely to be easily synthesized by recent developed selective chemical etching of transition-metal carbides (MAX phase). 
This will pave the way to tremendous applications of 2D TIs, such as ``ideal" conducting wire, multifunctional 
spintronic device, and the realization of topological superconductivity and Majorana modes for quantum computing.

\end{abstract}

\maketitle

\section{Introduction} \label{introduction}
The field of topological insulator (TI) started from the theoretical proposal of two-dimensional (2D) TI state 
in graphene.~\cite{kane2005PRL1, kane2005_PRL2, bernevig_quantum_2006} Though the band gap of graphene is too tiny to be observed,~\cite{Yao}
the conceptual achievement in the band topology has opened the door to the field of topological quantum states (TQSs).~\cite{TIreview, TIreview-2, MRS_weng:9383312}
The theoretical proposal~\cite{bernevig_science_quantum_2006} and experimental verification~\cite{konig_quantum_2007} of 2D TI in quantum well of HgTe/CdTe
have boosted the quick rising of the field of TIs. The idea of band topology has been extended to 3D system~\cite{3dTI1, 3dTI2, 3dTI3} and other
symmetry protected TQSs.~\cite{SPT2015} Recently, the band topology in metals, including the topological
Dirac semimetal,~\cite{Na3Bi, Cd3As2, Na3Biexp, Cd3As2exp} Weyl semimetal~\cite{wan,HgCrSe,TaAs_Weng, TaAs_arc, TaAs_node, TaAs_anomaly} and Node-line semimetal,~\cite{burkov, allcarbon_nodeLine2014, Cu3NPd, Cu3NPdKane} has also been intensively studied. Many of the material realization
of these TQSs are firstly predicted by theoretical calculations and then confirmed by experimental observations.~\cite{zhang_topological_2013, MRS_weng:9383312} 
The bulk-boundary correspondence of the topological matters is well known now and it is one of the most unique properties of them. 
For example, 2D TI is expected to host quantum spin Hall effect (QSHE) with 1D helical edge states, namely the electrons in such edge states 
have opposite velocities in opposite spin channels. Thus, the backscattering is prohibited as long as the perturbation does not break 
the time-reversal symmetry (TRS). Such helical edge states
are expected to serve as ``two-lane highway" for dissipationless electron transport, which promises great potential application in low-power
and multi-functional spintronic devices. Large band gap 2D TI is also crucial to realize the long-sought-for topological superconductivity and 
Majorana modes through proximity effect.~\cite{TS_PRL2008,TIreview-2} In this point of view, 2D TI is more preferred than 3D one, where 
the backscattering in the surface states is not fully prohibited. 

Compared with the number of well characterized 3D TI materials, fewer 2D TIs have been experimentally discovered.~\cite{ando_topological_2013, MRS_weng:9383312} 
The quantum wells of HgTe/CdTe~\cite{konig_quantum_2007} and InAs/GaSb~\cite{PhysRevLett.107.136603} are among the well-known
experimentally confirmed 2D TIs. 
Both of them require precisely controlled MBE growth and operate at ultra-low temperature. These experimental conditions make further studies 
hard and reduces the possible applications. There have been many efforts to find ``good" 2D TIs, which are expected to have the 
following advantages: (1) being easy to 
be prepared; (2) having large bulk band gap to be operated under room temperature or higher; (3) being chemically stable upon exposure to air; 
(4) being composed of cheap and nontoxic elements. The theoretical attempts for predicting good TIs can be roughly classified into two 
categories. 1) tuning the strength of spin-orbit coupling (SOC), i.e., the band gap, based on graphene-like honeycomb lattice, such as 
the low-buckled silicene~\cite{liu_quantum_2011}, 
chemically decorated single layer honeycomb lattice of Sn,~\cite{PhysRevLett.111.136804} Ge~\cite{PhysRevB.89.115429} and Bi or 
Sb,~\cite{BiX_Yao2014} and bucked square lattice BiF~\cite{XiangSQ}. 2) examining new 2D systems, which might be exfoliated from the 3D layered 
structural materials, such as ZrTe$_5$, HfTe$_5$~\cite{weng_transition-metal_2014} and Bi$_4$Br$_4$.~\cite{zhou_large-gap_2014}
Transition-metal dichalcogenide (TMD) in 1T'~\cite{qian2014quantum} and square-octagon haeckelite~\cite{ smnie2015, yanbh, daiying} structure
also belong to the later category. None of the above has been confirmed by experiments yet, though ZrTe$_5$, HfTe$_5$ and Bi$_4$Br$_4$ seem 
to be very promising, since they do exist experimentally and and their single layers are TIs without any additional tuning.

Regarding oxide materials, non of them is known to be TI in the experiment, though there are several theoretical proposals available in the literatures, 
e. g., 2D TI in single layer of iridate Na$_2$IrO$_3$,~\cite{PhysRevLett.102.256403} topological Mott insulator~\cite{227MottTI} and Weyl semimetal in pychorelcore $A_2$Ir$_2$O$_7$,~\cite{wan} axion insulator in spinnel Osmate~\cite{wanOs} and 3D TI in perovskite of YBiO$_3$~\cite{YBiO3} and heavily doped BaBiO$_3$.~\cite{BaBiO3} It is generally believed that the strong electronegativity of oxygen leads to
full ionization of cations and results in ionic bonds with large band gap, which makes
band inversion difficult. However, the noticeable advantages of oxygen compounds, i.e., naturally antioxidant and stable upon exposure to air, 
have stimulated continuous efforts in searching for new oxide TIs. 

In this paper, by using first-principles calculations we demonstrate that the functionalized MXenes~\cite{MXene2011, MXene2012, MXene2014} with oxygen,
M$_2$CO$_2$ with M=W, Mo and Cr, are 2D TIs. Our phonon calculations indicate that the crystal structures are dynamically stable. 
The band inversion, which is crucial to the nontrivial band topology, is found
to occur among the bonding and anti-bonding states of M $d$-orbitals. The results are robust against the use of different exchange-correlation functional approximations. The bulk band gap of W$_2$CO$_2$ is as large as 0.194 eV within generalized gradient approximation (GGA) and is enhanced to 0.472 eV 
within hybrid functional (HSE06).~\cite{heyd2003hybrid, heyd2006hybrid} Its $Z_2$ invariant is 1 and has conducting helical edge states. Recently, 
the 2D material MXene have been successfully obtained by the selective chemical etching of MAX phases --- M$_{n+1}$AX$_n$ ($n$=1, 2, 3 , ...), 
where M, A and X are a transition metal, an element of group 12-14, C or N, respectively.~\cite{MAX2000} The bare surfaces of MXene sheet are 
chemically active and are usually terminated by some atoms or chemical groups depending on the synthesis process, which are usually 
fluorine (F), oxygen (O), or hydroxyl (OH).~\cite{func_term_NMR, khazaei2013, khazaei2014} Therefore, we believe that our proposed M$_2$CO$_2$ 
can be probably realized experimentally in the future and thus, will advance the application of TI greatly. 

\section{Computational Details} \label{method}
First-principles calculations were carried out by using the
Vienna $ab~ initio$ simulation package (VASP)\cite{kresse1996_1,kresse1996_2}.
Exchange-correlation potential was treated within the generalized gradient
approximation (GGA) of Perdew-Burke-Ernzerhof type.\cite{Perdew1996} SOC
was taken into account self-consistently. The cut-off energy for plane wave expansion 
is 500 eV and the k-point sampling grid in the self-consistent process was 12$\times$12$\times$1. 
The crystal structures have been fully relaxed until the residual forces on each atom 
becoms less than 0.001 eV/\AA. A vacuum of 20 \AA ~between layers was considered 
in order to minimize the interactions between the layer with its periodic images.
PHONOPY has been employed to calculate the phonon dispersion \cite{togo2008}.
Considering the possible underestimation of band gap within GGA, non-local Heyd-Scuseria-Ernzerhof (HSE06)
hybrid functional\cite{heyd2003hybrid, heyd2006hybrid} is further supplemented
to check the band topology. To explore the edge states, we apply 
the Green's function method~\cite{MRS_weng:9383312} based on the tight-binding model with 
the maximally localized Wannier functions (MLWF)~\cite{marzari1997,souza2001} 
of $d$ orbitals of M and $p$ orbitals of C and O as basis set. MLWF are generated by 
using the software package OpenMX.~\cite{openmx, weng_mlwf}

 \begin{table}
\caption{
The total energies (in eV per unit cell) for the six possible configurations of W$_2$CO$_2$, Mo$_2$CO$_2$ and Cr$_2$CO$_2$. For each 
configuration, the structure is fully relaxed. In each unit cell, two oxygen atoms are required for full surface saturations. 
The symbols T, A, and B indicate three different absorption sites of oxygen atoms as depicted in Fig.~\ref{crystructure}.  All calculations
here are performed without spin polarization and within GGA.
}\label{stablestructure}
\begin{tabular*}{0.5\textwidth}{@{\extracolsep{\fill}}c|c|c|c}
 	\hline\hline
   sites of Oxygen & W$_2$CO$_2$ & Mo$_2$CO$_2$ & Cr$_2$CO$_2$  \\
    \hline
TT & 3.162 & 2.779 & 1.790 \\
 	\hline
AA & 1.189 & 1.112 & 0.618 \\
 	\hline
TB & 1.828 & 1.782 & 1.370 \\
 	\hline
TA  & 2.024 & 1.733 & 1.034 \\
         \hline
{\bf BB}  & {\bf 0.0} & {\bf 0.0} & {\bf 0.0} \\
         \hline
BA  & 0.878 & 0.698 & 0.329 \\
 	\hline\hline
\end{tabular*}
\end{table}
 
\begin{figure}
\includegraphics[scale=0.7]{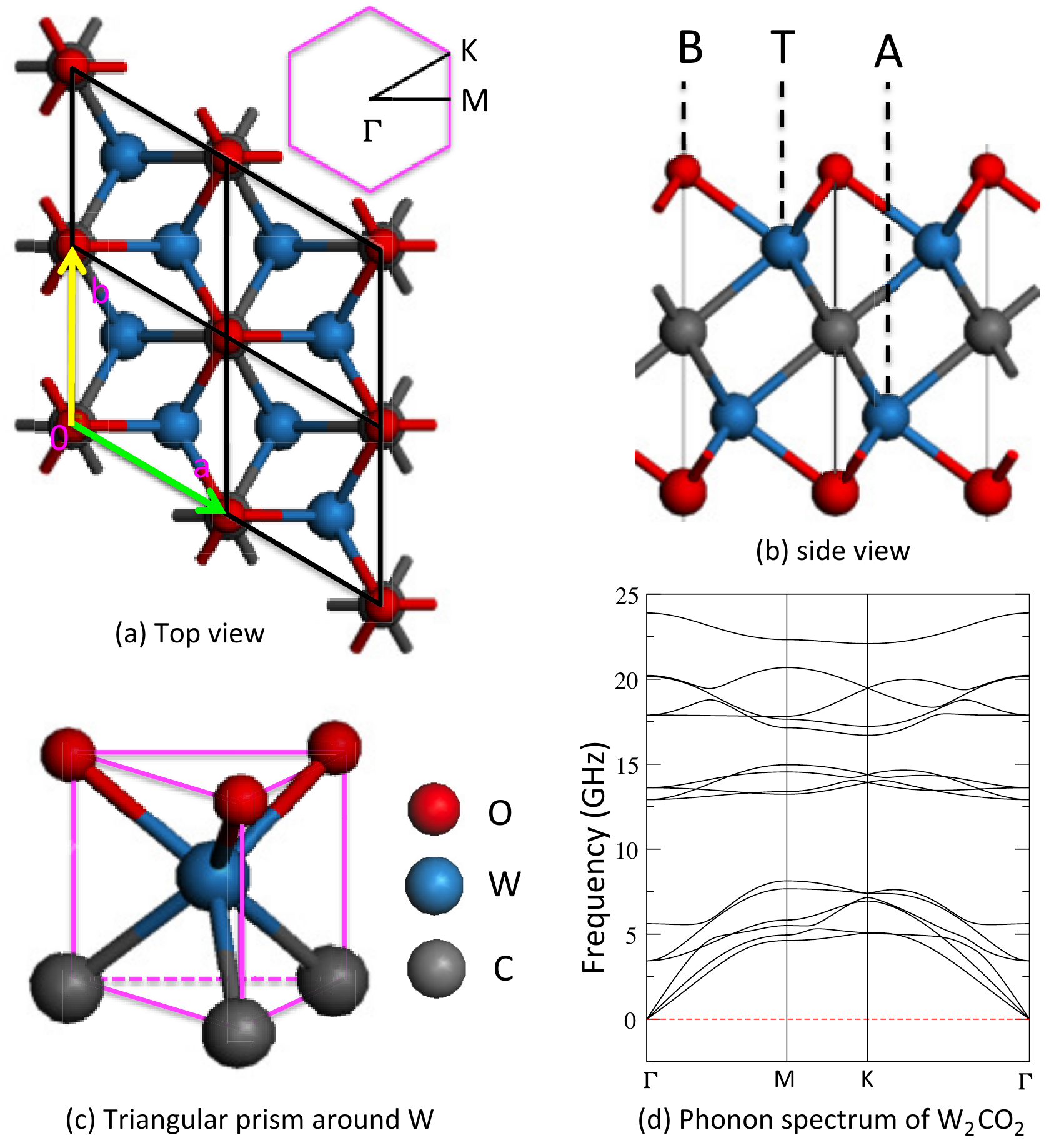}
\caption{(Color online) (a) Top view and (b) side view of optimized crystal structure of W$_2$CO$_2$ and its 2D Brillouin zone. 
There are three possible sites for oxygen decoration on either side of the surface, namely B (on top of C site), T (on top of W in top surface), A (on top of W in bottom surface). (c) The triangular prism formed by O and C ions surrounds W. (d) The phonon spectrum for optimized W$_2$CO$_2$.}
\label{crystructure}
\end{figure}

\section{Results and Discussion} \label{Results}
The crystal structure of oxygen functionalized MXene is shown in Fig. 1. Bare MXene M$_2$C is basically a three-layer structure with trigonal lattice. 
The C atoms form a layer, which is sandwiched between two M layers. The in-plane sites of M atoms
are (1/3, 2/3) and (2/3, 1/3), respectively, of the trigonal lattice of C atoms. The bare surfaces of MXene sheet are basically terminated
by M atoms and they are chemically reactive. Usually, the surfaces are terminated by F, O or OH depending on the synthesis process. This brings
a chance to tune the electronic properties of MXene by appropriate surface functionalization.~\cite{khazaei2013, khazaei2014} 
It has previously been shown that the MXenes with the full surface functionalizations are thermodynamically more favorable than 
the partial functionalizations,~\cite{khazaei2013} where the full surface functionalization requires two chemical groups per cell. 
As shown in Fig. 1, the oxygen atom
might occupy three different sites on the surface, namely A, B and T on each surface. Therefore, there are totally six combinations for decoration of two 
surfaces as listed in Table.~\ref{stablestructure}. For each case, the crystal structure is fully relaxed and the results of total energy calculations are 
summarized in Table.~\ref{stablestructure}. It is observed that the MXeen with BB-type oxygen functionalization obtain the lowest energy. In this
configuration the M atom sits inside of a trigonal prism formed by the surrounding C and O, which is very similar to Mo atom in 1H 
structure of MoS$_2$. The phonon spectrum of the energetically stable crystal structure is calculated and shown in Fig.~\ref{crystructure}(d). 
Obviously, there is no imaginary frequency, which means such structures are also dynamically stable. 

\begin{figure}
\includegraphics[scale=1.2]{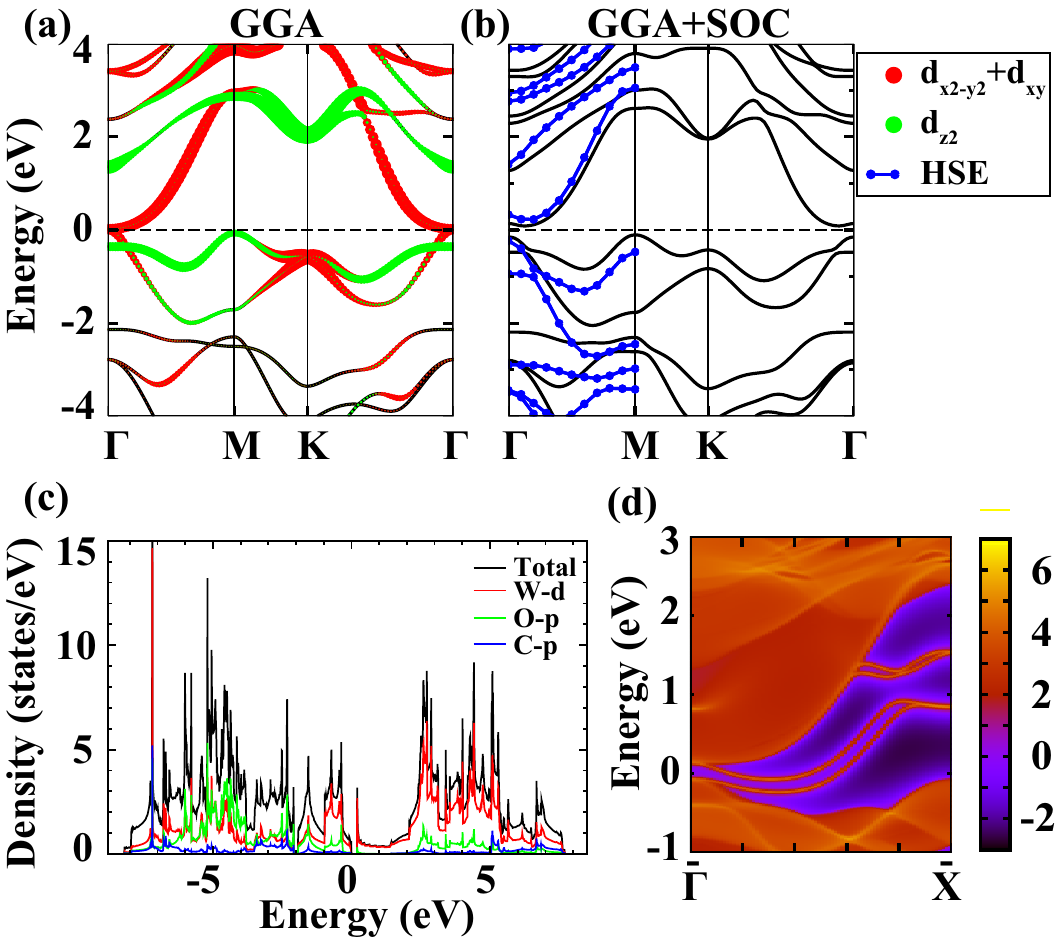}
\caption{(Color online) Band structure for W$_2$CO$_2$ calculated (a) without and (b) with spin-orbit coupling (SOC). The fat-band are scaled with the projected 
weight of different atomic orbitals within the eigenstates as shown in (a). The comparison with the bands from hybrid functional (HSE06) calculation including 
SOC is shown in (b). Total and projected partial density of states are shown in (c). The edge states along lattice constant $a$ is shown in (d). 
}
\label{bands2}
\end{figure}

The electronic band structure of W$_2$CO$_2$ is shown in Fig.~\ref{bands2}. It is found that there is a degenerate band touch point on the Fermi 
level when SOC is not considered. These degenerate states are mostly composed of $d_{x^2-y^2}$ and $d_{xy}$ 
orbitals of W atoms as shown by the fat-band analysis. If SOC is further considered, it becomes an insulator. Since it has inversion 
symmetry, the parity configuration of occupied bands at four time-reversal invariant momenta in the 2D BZ can be easily obtained 
and the $Z_2$ invariant is found to be 1. This indicates that W$_2$CO$_2$ is a 2D topological insulator with an indirect band 
gap as large as 0.194 eV. The topologically protected conducting edge states have Dirac cone like dispersion and connect the bulk 
valence and conduction bands.

\begin{figure}
\includegraphics[scale=1.0]{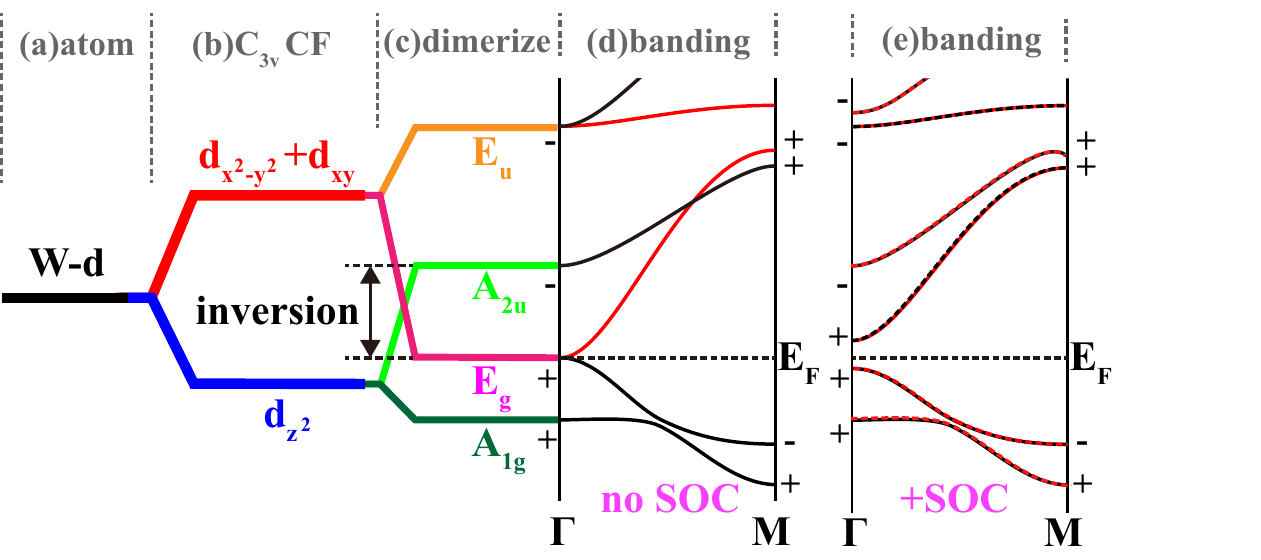}
\caption{(Color online) The band inversion mechanism in W$_2$CO$_2$. (a) W 5$d$ orbitals are split (b) under $C_{3v}$ crystal field (CF)
with double degenerate $d_{x^2-y^2}$+$d_{xy}$ and single degenerate $d_{z^2}$ orbitals around Fermi level E$_F$. (c) The dimerization
of two W atoms in each unit cell leads to bonding and anti-bonding states and band inversion between states with different parities (labeled by
+ and -). (d) The band
dispersion along $\Gamma$-M does not induce other band inversion. (e) Including SOC opens band gap and introduces spin degeneracy in each 
band, with dashed and solid line representing opposite spin channels.
}
\label{bandinv}
\end{figure}

The total and projected partial density of states, as well as the fat-band plot, clearly show that around Fermi level the W $d$ orbitals are 
dominant around the Fermi level. The $p$ orbitals of C and O are mainly located -2 eV below the Fermi level. The five $d$ orbitals 
of W atom are within the triangular prism crystal 
field of $C_{3v}$ symmetry. The $d_{xz}$ and $d_{yz}$ orbitals are double degenerate and are higher in energy due to the strong hybridization
with the $p$ orbitals of C and O. The $d_{x^2-y^2}$ and $d_{xy}$ orbitals are also degenerate and they are around the Fermi level. 
The $d_{z^2}$ orbital has quite weak hybridization with ligand elements since it points to the center of the ligand triangle. 
It is also around the Fermi level and slightly lower than $d_{x^2-y^2}$+$d_{xy}$. Therefore,  in order to uncover the low-energy physics
of W$_2$CO$_2$, we just need to take into account three three $d$ orbitals,
$d_{x^2-y^2}$, $d_{xy}$ and $d_{z^2}$, of the W atoms. It is noticeable that there are two W atoms in one unit cell. As shown in Fig.~\ref{bandinv}, 
the above selected $d$ atomic orbitals form bonding and anti-bonding states. The bonding and anti-bonding
states of $d_{z^2}$ orbitals are even ($A_{1g}$) and odd ($A_{2u}$), respectively. Similarly, those from $d_{x^2-y^2}$+$d_{xy}$ are
even ($E_g$) and odd ($E_u$) but having double degeneracy. The band inversion happens between the double degenerate $E_g$ 
and non-degenerate $A_{2u}$ states,which brings the nontrivial topology of bands. When the banding effect is considered from 
$\Gamma$ to M, the double degenerate $E_{g}$ states are split and there is no other band inversion happens. 
Further, the SOC introduces the spin degree of freedom and it opens a gap around
Fermi level as shown in Fig.~\ref{crystructure}. Due to the heavy W element, the band gap is found to be as 
large as 0.194 eV. Considering the possible underestimation of the band gaps within GGA, the hybrid functional (HSE06) calculation is used
to check whether the above band inversion around $\Gamma$ is robust. It is found that the band inversion is kept and the band 
gap is enhanced to as large as 0.472 eV. Such large band gap 2D TI has an advantage in observing quantum spin Hall effect at room-temperature 
or higher, which is appropriate for device applications.

\begin{figure}
\includegraphics[scale=0.5]{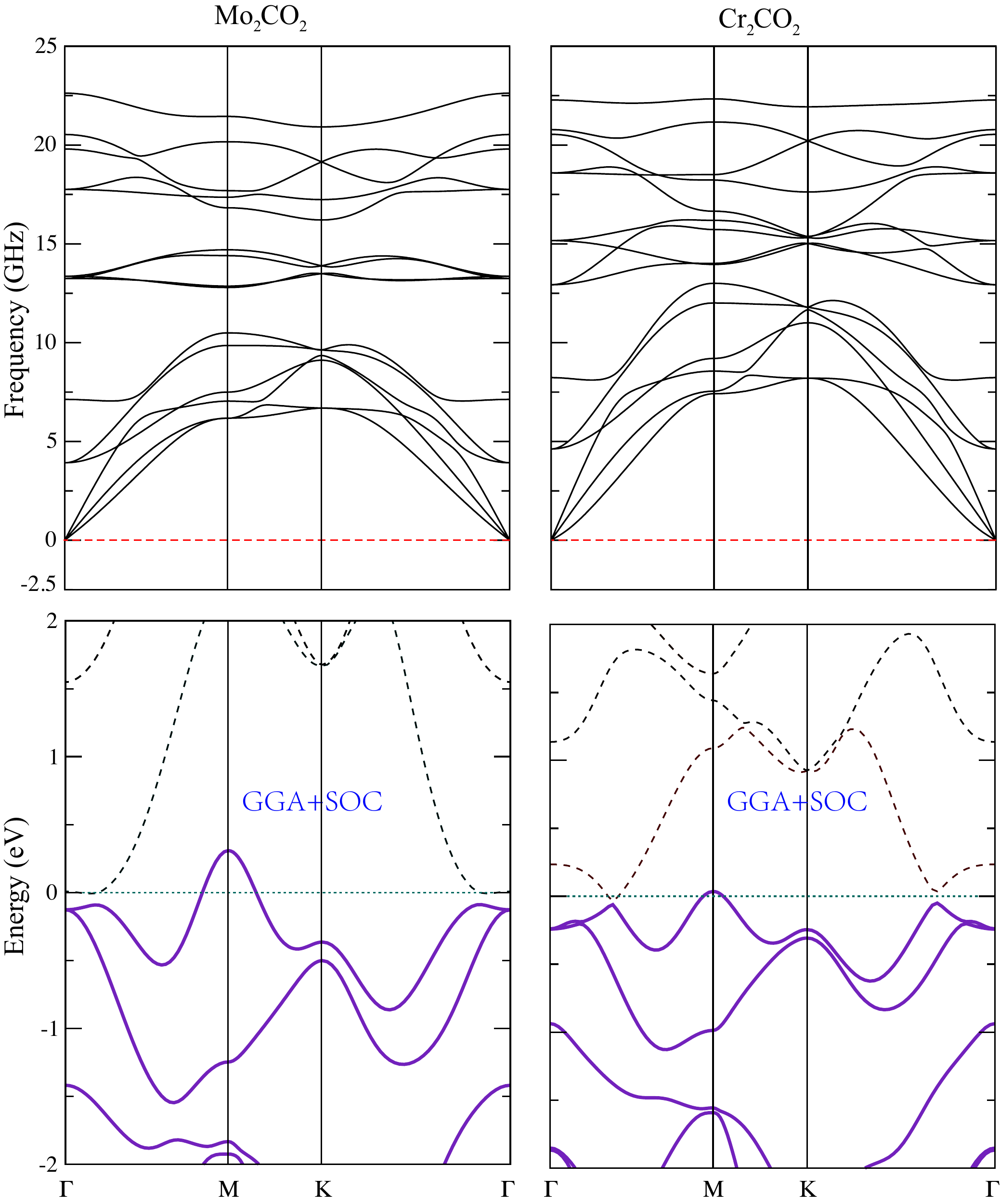}
\caption{(Color online) The phonon spectrum (upper panel) and band structure with SOC (lower panel) for optimized stable Mo$_2$CO$_2$ (left panel)
and Cr$_2$CO$_2$ (right panel) The assumed occupied (unoccupied) bands are denoted with solid (dashed) lines. 
}
\label{mocr}
\end{figure}

For Mo$_2$CO$_2$ and Cr$_2$CO$_2$, both of them have the similar crystal structure as W$_2$CO$_2$ and are also 
dynamically stable as seen from the phonon spectra shown in Fig.~\ref{mocr}. Since Mo and Cr have weaker SOC than W, they are expected to have 
narrower band gap than W$_2$CO$_2$. In fact, both of them show band structure of semimetals by having compensated electron and hole Fermi pockets. 
However, at each k point, one can still find well defined band gap with a curved Fermi level. The assumed occupied bands, as denoted with solid lines
in Fig.~\ref{mocr}, have the same $Z_2$ number as W$_2$CO$_2$. Therefore, both of them share the same band topology, as well as the 
underlying physics, with W$_2$CO$_2$. 

Furthermore, we have investigated the correlation effect in $d$ electrons of transition-metal M. 
For quite delocalized 5$d$ electrons in W and 4$d$ in Mo, detailed GGA+$U$ ($U$ is the parameter for onsite Coulomb interaction) 
calculations ($U<$4.0 eV) show that the correlation effect is negligible and the ground state of W$_2$CO$_2$ and M$_2$CO$_2$ are 
always nonmagnetic. The results discussed above are robust. However, for 3$d$ transition-metal Cr, simple GGA+$U$ ($U>$2.0 eV) 
calculation for BB configuration gives out solution with magnetic properties. The above non-spin polarization calculations for 
Cr$_2$CO$_2$ are helpful to understanding the band gap dependence on the atomic number in that column, but not realistic for 
Cr$_2$CO$_2$ itself. However, this gives us the spin degree of freedom in material design based on MXene, which is very crucial
to seeking Chern insulators hosting quantum anomalous Hall effect.~\cite{Yu02072010qahe, Chang12042013qahe} The 
detailed studies on the dependence of $U$ value and magnetic configurations are left for future work.

\section{Conclusions}
 Based on the first-principles calculations, we have predicted that a family of oxygen functionalized MXene M$_2$CO$_2$ (M=W, Mo and Cr) are 2D TIs.
 The representative W$_2$CO$_2$ has robust band inversion, nontrivial $Z_2$ invariant and large band gap of 0.194 (0.472) eV within GGA (HSE06). It 
 might satisfy all the four criteria of a "good" 2D TI. (1) easier production process by selective chemical etching method; (2) hosting QSHE at ambient condition; (3) high stability and antioxidant upon exposure to air; and (4) low production cost and consisting of environmental friendly elements. 

Inspired by these findings, one can naturally think about obtaining more 2D TI candidates in other functionalized MXenes. 
The large number of MAX (more than 60)~\cite{MAX2000, C2NMXene} brings many possibilities of MXene~\cite{MXene2014} 
and huge space in finding topologically nontrivial materials. The possible changes would include replacing O$^{2-}$ 
with F$^{-}$ or (OH)$^{-}$,~\cite{khazaei2013, khazaei2014, tiDirac} varying transition-metal M, replacing C with N or B,~\cite{C2NMXene} 
tuning the number of layers $n$ in MXene M$_{n+1}$C$_n$, etc. The material design or property tailoring with a single or any 
combination of the above changes will lead to more and better 2D TIs or Chern Insulators.


\section{Acknowledgments}
H.W., Z.F. and X.D. acknowledge the supports from National Natural Science Foundation of China (Grant Nos. 11274359 and 11422428), 
the National 973 program of China (Grant Nos. 2011CBA00108 and 2013CB921700) and the "Strategic Priority Research Program (B)" 
of the Chinese Academy of Sciences (Grant No. XDB07020100). H.W. thanks the hospitality during his
stay in Tohoku University and part of this work has been done there. Y.K. acknowledges to 
the Russian Megagrant project (Grant No. 14.B25.31.0030). Both Y.L. and Y.K. are
supported by JST, CREST, "A mathematical challenge to a new phase of material sciences" (2008-2013). 
Partial of the calculations were preformed on TianHe-1(A), the National Supercomputer Center in Tianjin, China.

\bibliographystyle{unsrt}
\bibliography{MXene_ref}
\end{document}